\documentclass{nature}

\usepackage{float}
\usepackage{amsmath}
\usepackage[labelfont=bf]{caption}
\usepackage{graphicx}
\makeatletter
\let\saved@includegraphics\includegraphics
\AtBeginDocument{\let\includegraphics\saved@includegraphics}
\renewenvironment{figure}{\@float{figure}}
{\end@float}
\makeatother
\usepackage{xcolor}
\usepackage{amsmath}
\usepackage{comment}
\usepackage[%
]{hyperref}

\newcommand{\trR}{$\Delta R/R$}

\newcommand{\Tc}{$T_{c}$}

\newcommand{\PFE}{$P_{FE}$}

\newcommand{\chithree}{$\chi_{3fold}^{\parallel}$}

\title{Non adiabatic dynamics of the ferroelectric soft mode}

\author{Gili Scharf$^1$, Lara Donval$^1$, Leah Ben Gur$^1$, Alon Ron$^1$}

\begin{document}

\maketitle

\begin{affiliations}

 \item Raymond and Beverly Sackler School of Physics and Astronomy, Tel-Aviv 69978, Israel
\end{affiliations}

\begin{abstract}

Most microscopic descriptions of structural dynamics assume the Born-Oppenheimer separation, where electrons adjust adiabatically to ionic motion. When this separation breaks down, electronic and lattice degrees of freedom can evolve on different timescales, giving rise to new physical phenomena beyond the adiabatic limit. Here we use time-resolved, phase-sensitive second-harmonic generation and pump-probe reflectivity to reshape the ferroelectric free-energy landscape of SnTe while separately tracking polar order and coherent lattice motion. When photoexcitation transiently suppresses the double-well barrier, polarization dynamics become strongly nonlinear, while the coherent phonon dynamics remain harmonic. This decoupling cannot be described by a single adiabatic coordinate for the electronic polarization and ionic positions. We provide a unifying physical description for the non adiabatic dynamics of the ferroelectric mode and the mixed displacive/order-disorder nature of SnTe based on a separation of scales for the renormalization of the ferroelectric stiffness.

\end{abstract}

Structural dynamics in solids are commonly described within a Born-Oppenheimer framework, where ions evolve on an electronic potential energy surface and the electronic polarization is assumed to follow the lattice coordinate adiabatically \cite{BornOppenheimer1927}. This viewpoint underlies much of the standard microscopic modeling of phonons and structural instabilities, including first principles calculations \cite{HohenbergKohn1964,KohnSham1965,GonzeLee1997,Baroni2001}.

Breakdowns of the adiabatic Born-Oppenheimer separation are well established when electronic relaxation competes with lattice motion, producing non adiabatic phonon renormalization and shaping coherent phonon generation in pump–probe experiments \cite{zeiger1992theory,LazzeriMauri2006,Saitta2008}. Ferroelectrics provide an especially sharp arena because the order parameter itself is an electronic polarization, and equilibrium spectroscopy has long indicated an additional slow central-mode polarization channel distinct from the THz soft phonon\cite{Petzelt1984}. Additionally, time-resolved measurements have shown that the DC-like strain and ferroelectric polarization can decouple\cite{Hoang2025}.  Yet direct tests of whether the polarization remains dynamically locked to the coherent soft-mode coordinate remain unexplored.

To search for non-adiabaticity of the ferroelectric mode, we choose SnTe, a carrier tunable ferroelectric with a shallow double-well \cite{Pawley1966,Littlewood1980,SugaiMurase1982,Polking2012} and evidence for local inversion breaking above the macroscopic transition \cite{Mitrofanov2014PRB,Fornasini2021JPCM,Aggarwal2016JMat,Chassot2024NanoLett}. 
We combine time resolved second harmonic generation (SHG), which yields the magnitude and sign of the ferroelectric polarization,\PFE, with ultrafast pump-probe reflectivity tracking the coherent soft mode like lattice response. 
The magnitude and the sign of $P_{\mathrm{FE}}$ in our SHG measurements are provided through interference between multiple nonlinear processes originating from the sample \cite{seyler2022direct}. 
By measuring $P_{\mathrm{FE}}(t)$ and the coherent lattice coordinate simultaneously, we directly test whether polarization and ionic motion remain adiabatically locked under strong photoexcitation. 
By optically renormalizing the double-well curvature through photoexcited carriers, we drive ultrafast polarization switching while the coherent oscillation frequency remains nearly unchanged at high excitation density, demonstrating a dynamical separation between order parameter motion and the lattice coordinate. 
These results expose the hybrid electronic-ionic character of the ferroelectric mode in SnTe.

\par


\begin{figure}[!t]
    \centering
    \includegraphics[width=\linewidth]{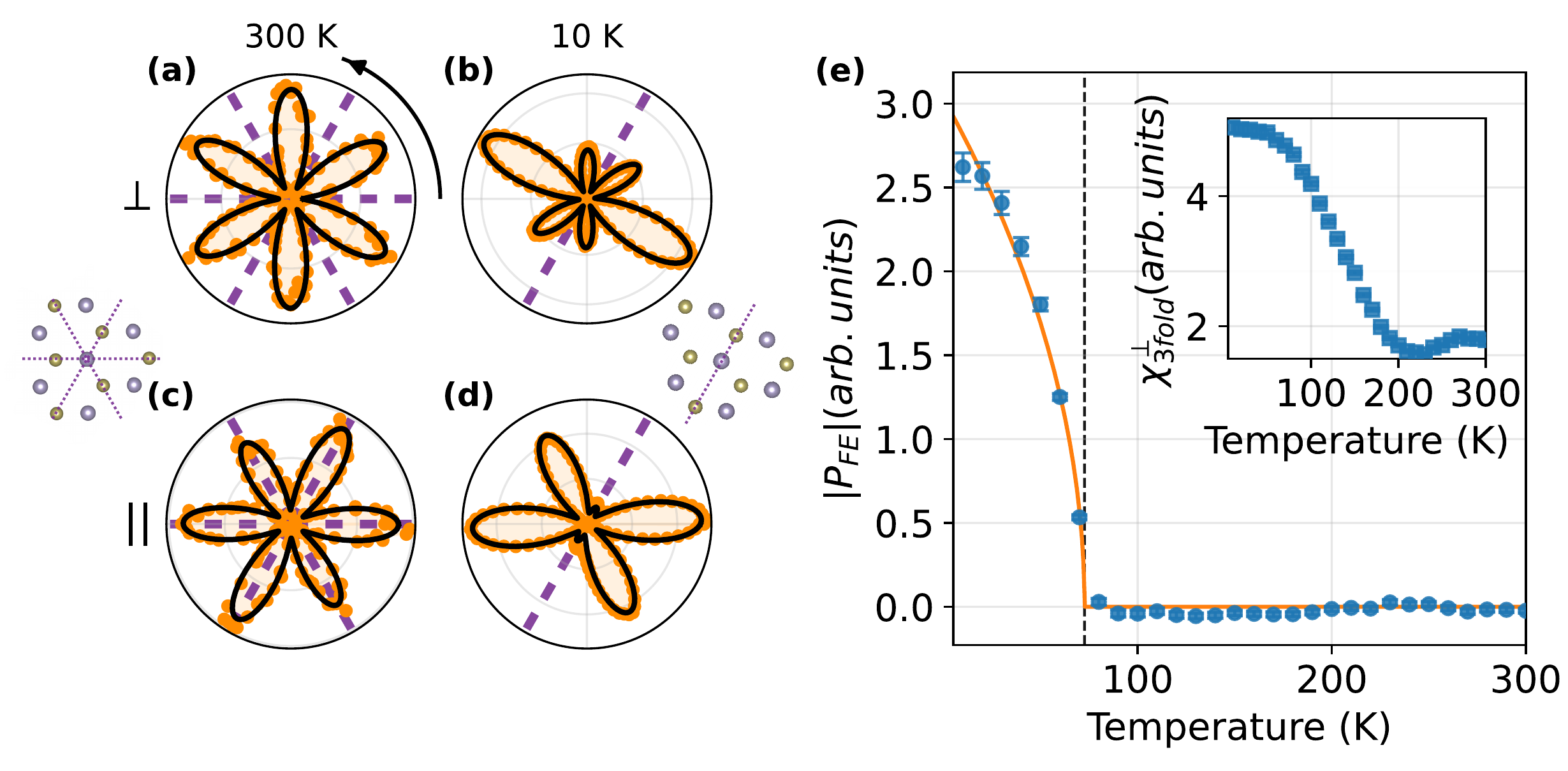}
    \vspace{-3.5em}
    \caption{(a-d) The SHG intensity as a function of rotation angle of the SnTe (111) surface taken at temperatures of 300K (a)+(c) and 10K (b)+(d) in the parallel (c,d) and perpendicular (a,b) configurations. Orange circles are the datapoints and the solid black line is a fit to the data. Dashed purple lines mark the direction of mirror planes. Illustrations of the SnTe (111) surface above and below the transition are shown at the sides, Sn and Te planes shifts are exaggerated for clarity, mirror planes are marked with dashed lines. Symmetry breaking is exaggerated below the transition. 
    (e) Magnitude of the ferroelectric order parameter \PFE, as extracted from the fits to the data. Blue circles are data points and the orange curve is a mean field fit. The dashed black line at 71K marks \Tc~as extracted from the fit. Inset shows $\chi_{3fold}^{\perp}$ as a function of the temperature.}
    \label{Figure 1}
\end{figure}

Below $T_c$, SnTe develops a macroscopic polarization along one of the eight equivalent $\langle 111\rangle$
directions due to a relative displacement of the Sn and Te planes, resulting in the loss of inversion symmetry\cite{iizumi1975phase}. We probe this transition by measuring the SHG signal as a function of temperature from the (111) surface with normal incidence geometry. Above $T_c$, SnTe is centrosymmetric, nonetheless, even at room temperature, the rotational anisotropy SHG (RA-SHG) signal is nonzero and sixfold symmetric due to a surface electric dipole contribution as shown in Figure \ref{Figure 1}(a,c). Illuminating at normal incidence, polarization directions that are $180^\circ$ relative to one another are equivalent and the threefold symmetry of the (111) surface (point group $3m$) manifests as a sixfold pattern in the RA-SHG. 
The signal also follows the three mirror planes depicted by the purple dashed lines corresponding to the mirror planes in the illustration of the surface to the left of panel (c). Figure \ref{Figure 1}(b,d) shows the SHG signal measured below the ferroelectric transition temperature $T_c$, clearly the threefold symmetry of the (111) surface and two of the mirror symmetries are broken by the ferroelectric transition. The distorted structure, along with the remaining mirror plane are illustrated next to panel (d). This symmetry reduction indicates that \PFE~ is not aligned with the surface normal of the measured (111) plane, but instead corresponds to a $\langle 111\rangle$ domain variant whose projection onto the (111) plane is nonzero. The expected RA-SHG under these conditions is:
$\begin{aligned}
    I_{\parallel}(\phi)&=\big(\chi_{3fold}^{\parallel}\,\cos 3\phi + \chi_{2fold}^{\parallel}\,\cos\phi\big)^2,\\
    I_{\perp}(\phi)&=\big(\chi_{3fold}^{\perp}\,\sin 3\phi - \chi_{2fold}^{\perp}\,\sin\phi\big)^2
\end{aligned}$
Here $\chi_{3fold}^{\parallel}$ and $\chi_{3fold}^{\perp}$ parameterize the noncritical background contribution (dominated by the surface electric dipole response) while $\chi_{2fold}^{\parallel}$ and $\chi_{2fold}^{\perp}$ capture the ferroelectric contribution for the parallel ($\parallel$) and perpendicular ($\perp$) polarization configurations (see supplementary material section 2 for the full derivation). We fit the data to these expressions, and the results are shown as solid lines overlaying the RA-SHG data. We plot $P_{\mathrm{FE}}\propto\chi_{2fold}^{\perp}$ as a function of temperature in Figure \ref{Figure 1}(e). $P_{\mathrm{FE}}$ is zero above $T_c$ and follows a typical mean-field behavior
below it. A fit to mean-field behavior results in a $T_c\sim$ 71K, of the order of the previously
reported value for SnTe with a hole density of $3.6\cdot 10^{20}[cm^{-3}]$ \cite{sugai1977carrier}. The inset to Figure \ref{Figure 1}(e) shows the magnitude of $\chi_{3fold}^{\perp}$ as a function of temperature. The signal grows monotonically above $T_c$ but does not exhibit the same mean-field like behavior as $P_{\mathrm{FE}}$. This behavior is typical of systems in which the lattice parameters evolve continuously as a function of the temperature \cite{ron2019dimensional}.

\par
To probe the soft mode associated with the ferroelectric phase transition, we employ pump-probe reflectivity. We use pump and probe pulses centered around 1400nm and 900nm, respectively. Figure \ref{Figure 2}(a) shows the transient relative change in the reflectivity \trR\ as a function of time delay between the pump and the probe pulses for various temperatures above and below \Tc. Above \Tc, the signal is composed only of exponentially decaying components, while below \Tc, coherent oscillations are observed. Investigation of the data shows strong softening of the oscillations as the temperature is increased towards \Tc, evident by the change in the peak positions until they become too weak to discern from the noise at \Tc. The curves are fit to the following expression: $\frac{\Delta R}{R}=A_1 e^{-t/\tau_1}+A_2 e^{-t/\tau_2}\cos(\omega t+\phi)+c$, where $\omega$ is the angular frequency of the phonon, $\tau_1$ and $\tau_2$ are the electronic and phononic lifetimes, and $c$ captures slow dynamics beyond our measurement range. $\omega^2(T)$ was extracted from the fit, and plotted as a function of the temperature in
Figure \ref{Figure 2}(b). Between $T_c$ and 30K, $\omega^2(T)$ shows a mean-field Cochran like behavior \cite{Cochran1960} extrapolating to $T_0\sim$ 100K. The fast Fourier transform (inset) of the data in (a) is shown for comparison and follows the same softening behavior. Finally, the extracted oscillation frequencies at selected temperatures agree with previously reported soft mode frequencies from inelastic scattering measurements of samples with a similar transition temperature, supporting the assignment of the coherent oscillations in \trR\ to the ferroelectric mode \cite{o2017inelastic}.

\begin{figure}[H]
    \centering
    \includegraphics[width=0.65\textwidth]{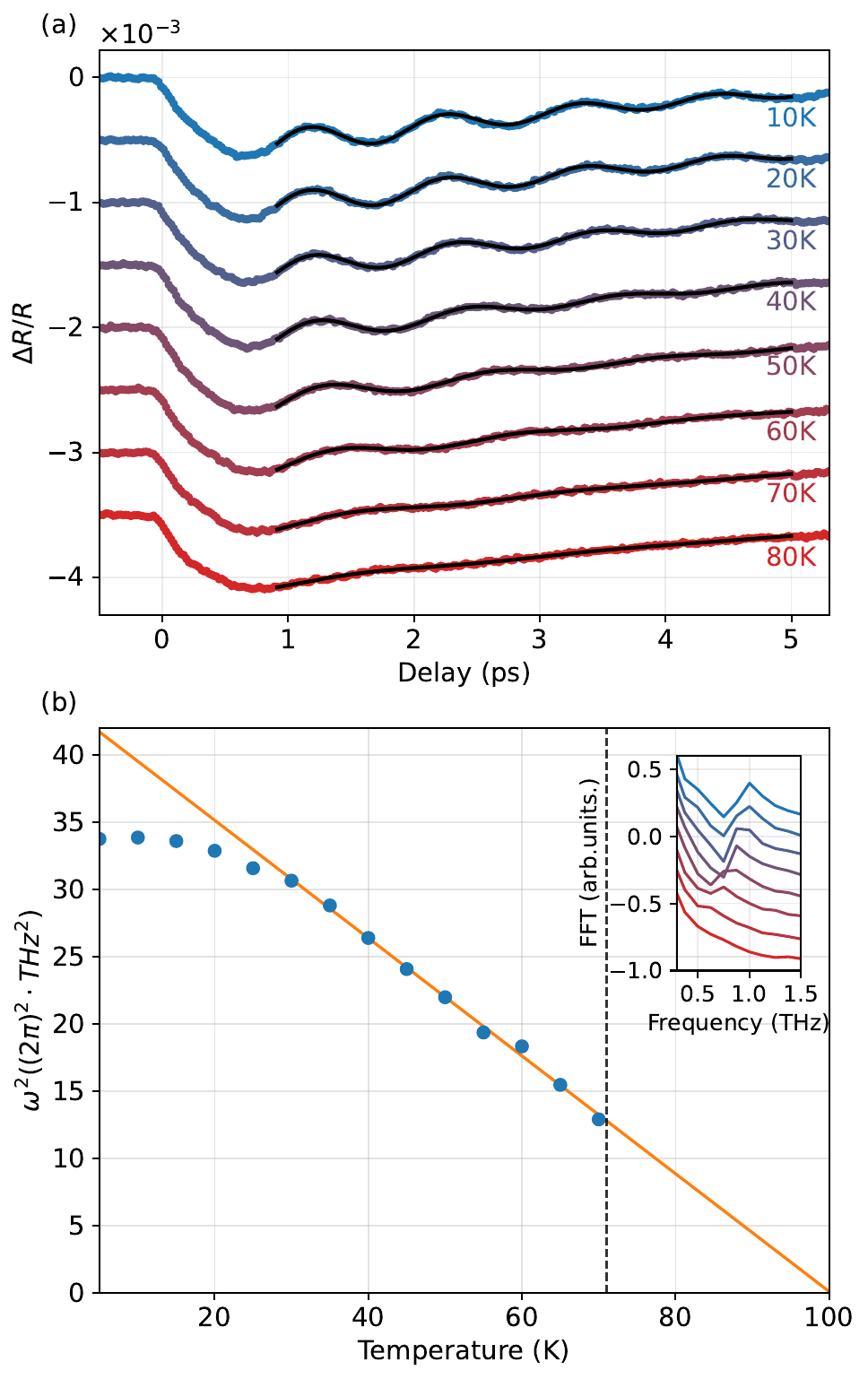}
    \vspace{-1em}
    \caption{(a) Pump probe reflectivity at a fluence of $\sim4.76 [mJ/cm^2]$ as a function of temperature from 10K to 80K, temperatures are marked below the data, black lines are fits to the data, different traces are vertically shifted for clarity. (b) $\omega^2$~extracted from fitting the data shown in (a) as a function of temperature. The orange line is a linear fit to the data. The dashed line marks \Tc~extracted from the SHG measurements. The inset shows the Fourier transform of the data in (a).}
    \label{Figure 2}
\end{figure}

\par

\begin{figure}[p]
    \centering
    \includegraphics[width=0.71\textwidth]{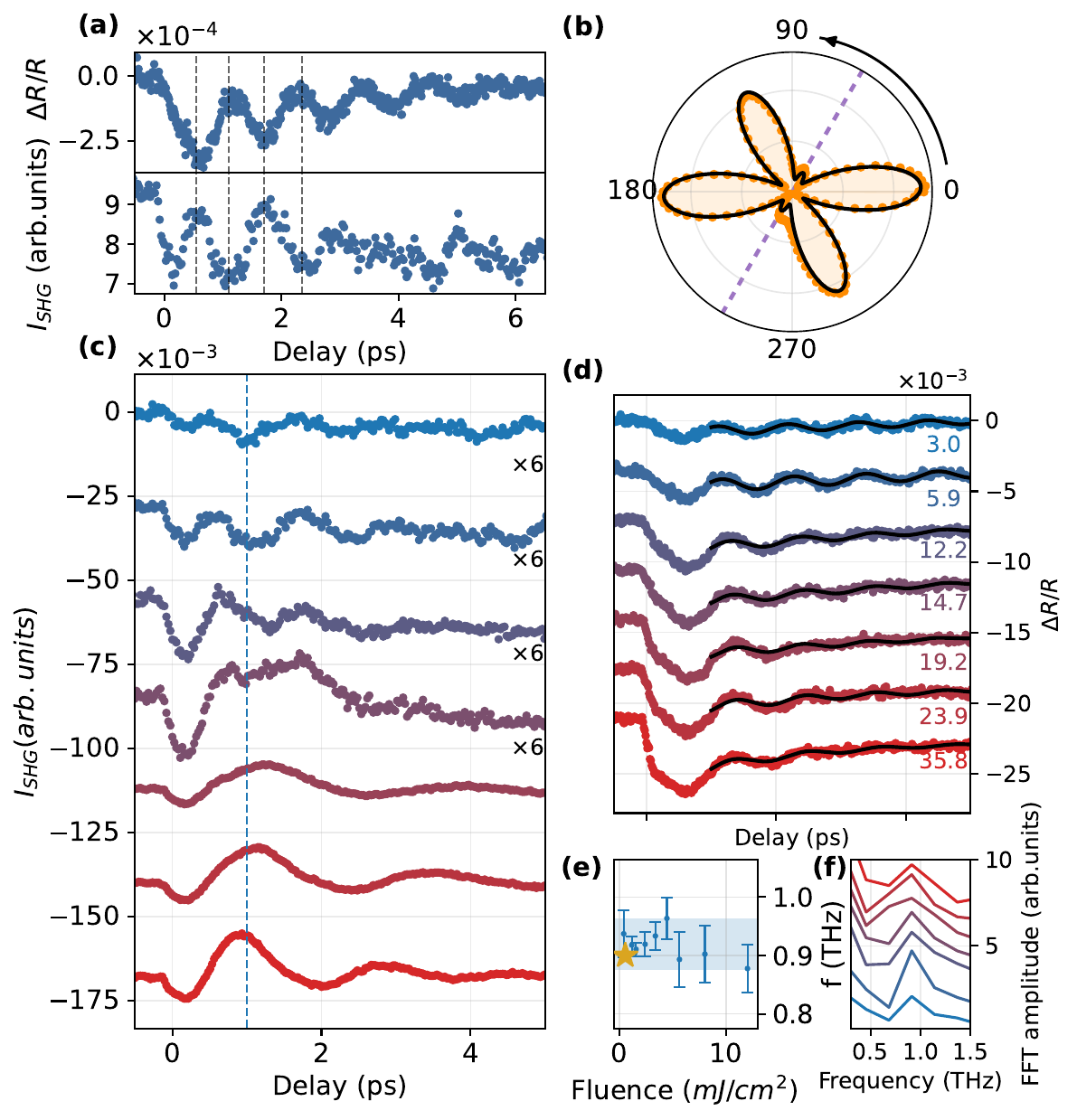}
    \vspace{-1em}
    \caption{(a) Time resolved SHG  (bottom) and \trR~(top) at a pump fluence of $\sim 5.9[mJ/cm^2]$~and $T=$5K. Dashed lines are guides to the eye for the joint oscillation minima and maxima positions. (b) RA-SHG pattern taken in the absence of the pump. Measurements in (a) and (c) were taken along the mirror plane at $60$\textdegree. A dashed purple line marks the mirror plane. (c) SHG intensity as a function of time delay for various pump fluences coded by color, similar to panel (d). All data is shifted for clarity. The top four traces are multiplied by a factor of 6 for clarity. (d) \trR~as a function of time delay for various fluences taken at 5K. Numbers next to data are the respective fluences in $mJ/cm^2$. Black solid lines are fits to the data. (e) Frequency extracted from the pump probe data in (d) as a function of the pump's fluence. (f) Fourier transform for the data in (d), colors mark the same fluences as in (d).}
    \label{Figure 3}
\end{figure}

Coherent phonons contribute to \trR\ through pump induced modulations of the reflectivity associated with the coherent lattice coordinate $Q$. To first order, the oscillatory component of $\Delta R$~can be written as $\Delta R_{osc} \simeq \left(\frac{dR}{dQ}\right)\Delta Q$~\cite{zeiger1992theory}, so when the soft mode is coherently excited, \trR\ acquires an oscillatory component at the mode frequency. In contrast, the dynamics of \PFE\ can be probed directly by time-resolved SHG. In our geometry, the ferroelectric twofold SHG field interferes with the sixfold background field, providing phase sensitivity that we use to determine the sign of \PFE. Time resolved SHG measurements at 5K were taken at a selected polarizer–analyzer angle along the low-temperature mirror plane and shown in Figure \ref{Figure 3}(b). In \ref{Figure 3}(a), the transient SHG measurement at low pump fluence ($5.9[mJ/cm^2]$) is compared to the \trR~measurement at the same fluence. Both signals oscillate at a frequency of $\sim 0.9~\mathrm{THz}$, indicating that they both originate from oscillations of the ferroelectric soft mode. However, as the fluence is increased, the SHG dynamics change dramatically as shown in Figure \ref{Figure 3}(c). The dip between the first two oscillatory peaks gradually fills and evolves into a broad, slow feature that spans $\sim 2[ps]$, nearly twice the period of a single low-fluence oscillation.  Additionally, as the fluence is increased, the magnitude of the slow oscillations becomes much larger than it was under low fluences (note the multiplicative factor of 6 used for plotting the low fluence data).
Strikingly, neither the slowdown nor the significant amplitude increase observed in SHG are present in \trR. As shown in Figure \ref{Figure 3}(d,e), the coherent oscillation frequency extracted from \trR~remains between $0.9$ and $0.95$ THz across the full fluence range. A Fourier transform corroborating the nearly constant frequency is shown in \ref{Figure 3}(f). Additionally, the dramatic magnitude increase observed in SHG, at the same fluence range where the slowdown occurs, is not seen in \trR. At low fluence, the SHG response and the oscillatory component of \trR~exhibit the same eigenfrequency and a fixed phase relationship, consistent with the two observables being dynamically locked. As the fluence is increased, this correspondence breaks down, SHG develops an additional slow component (and waveform distortion) while the reflectivity oscillation remains well described by a single damped mode with essentially unchanged frequency. We interpret this as an unlocking of the coherent modes sensed by \trR~and SHG under strong photoexcitation. Given that SHG provides a direct readout of \PFE, while the oscillatory reflectivity response is commonly treated as a proxy for the soft mode coordinate, the observed unlocking suggests that \PFE~is no longer adiabatically locked to a single lattice coordinate in the high fluence regime. This dynamical separation highlights the hybrid electronic and ionic character of the ferroelectric mode. 

\par

\begin{figure}[p]
    \centering\includegraphics[width=\linewidth,height=0.65\textheight,keepaspectratio]{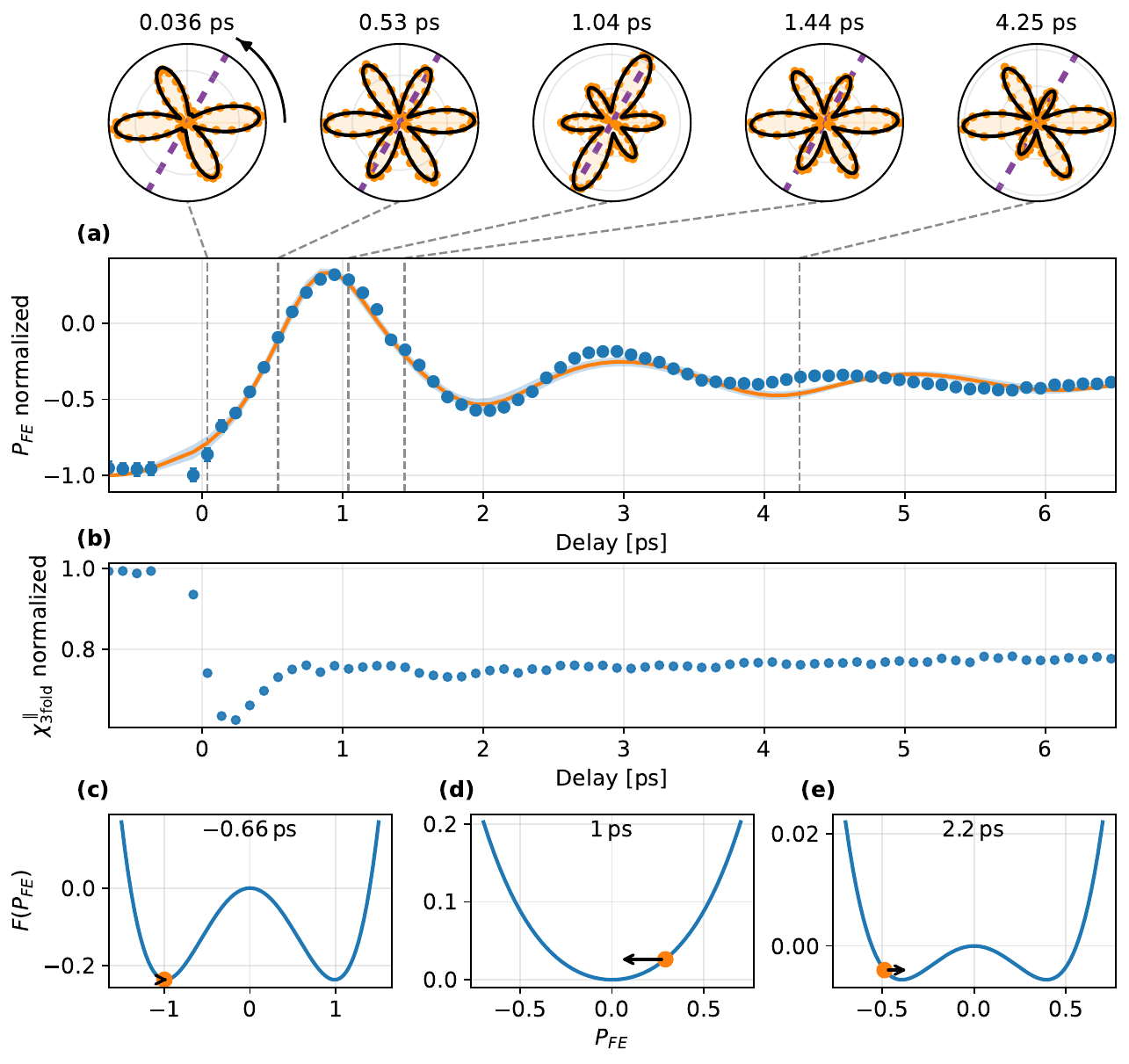}
    \vspace{-1em}
    \caption{(a) $P_{FE}$ Normalized to its equilibrium value as a function of pump delay time for a fluence of $\sim35.8 [mJ/cm^2]$~at 5K. Blue dots are the data, orange line is a fit and the blue shading is the fit $1\sigma$~confidence band. Dashed lines mark specific times for which normalized RA-SHG patterns are displayed above the plot. (b) \chithree~normalized to its equilibrium value as a function of time delay. (c-e) The double-well potential for the extracted fitting parameters at three specific delay times. Blue curve is the Landau potential, orange dot is the position value of $P_{FE}$~and the black arrow marks the direction and magnitude of the velocity. (c) Before quenching the barrier, (d) after quenching the barrier and (e) when it has partially recovered}
    \label{Figure 4}
\end{figure}

To better understand the origin of the slow oscillations, we chose a representative fluence where the slow oscillations are most pronounced ( $\simeq35.8[mJ/cm^2]$) to perform time resolved RA-SHG measurements. We collect a full RA-SHG  pattern for each time delay in the parallel polarizer analyzer configuration, where the slow oscillations were most pronounced. Fits to the RA-SHG patterns are shown in Figure \ref{Figure 4}(a) and are used to extract \PFE$(t)$~and \chithree$(t)$~which are plotted and normalized to their equilibrium values, as a function of time delay in Figure \ref{Figure 4}(a) and (b), respectively. The slow oscillations are very pronounced in \PFE~and are absent in \chithree.  There is a striking relation between the sign of \PFE~and the shape of the RA-SHG patterns shown in Figure \ref{Figure 4}. The patterns observed at times where \PFE$<0$~are elongated perpendicular to the mirror plane, whereas the ones which correspond to \PFE$>0$ are elongated parallel to it. The different directions correspond to the orientation of \PFE~as a function of time delay. In other words, our pump induced dynamics result in ultrafast switching of the sign of \PFE~along the same axis to which it was aligned before the arrival of the pump.  In the context of the Landau free energy double-well potential, our pump pulse induces ultrafast dynamics, which results in crossing between the two wells. Accordingly, when \PFE$ = 0$, the threefold symmetry of the (111) surface is momentarily restored.

\par
Similar order parameter dynamics were recently observed in several charge density wave systems\cite{huber2014coherent,yusupov2010coherent,schaefer2014collective} and understood as quenching of the double-well barrier as a result of the photoexcitation. Similarly, we introduce a time dependent double-well potential for the order parameter \PFE:
\begin{equation}
    V(P_{FE})=\frac{1}{2}(\eta(t)-1)P_{FE}^2+\frac{1}{4}P_{FE}^4
\end{equation}
Where $\eta(t)=\eta_0\cdot g(t)\cdot \exp(-t/\tau)+\eta_\infty$~is the time dependent renormalization of the quadratic coefficient (the oscillator stiffness). When $\eta(t)$~increases to one the double-well collapses. $\eta_0 \propto F$ (the fluence) is the excitation strength,  $\tau$ is the characteristic timescale associated with the relaxation of the double-well to its equilibrium value and $\eta_\infty$ symbolizes relaxation processes much slower than our measurement range. $g(t)=\mathrm{erf}(\sigma t)$ is the convolution of a step function with a Gaussian that models the response time of the sample to the optical excitation and is characterized by a timescale $1/\sigma$.  From this, one can describe the dynamics of \PFE~using the following equation of motion:
\begin{equation}
    \frac{1}{\omega^2}\frac{\partial ^2}{\partial t^2}P_{FE}-(\eta(t)-1)P_{FE}+P_{FE}^3+\frac{2\gamma}{\omega^2}\frac{\partial}{\partial t}P_{FE}=0
\end{equation}
Where $\omega$~ is the ferroelectric mode frequency at 5K, the temperature of the measurement, and $\gamma$~is a characteristic damping rate for \PFE.  A solid line in Figure \ref{Figure 4}(a) represents the fit of \PFE$(t)$~to this model. See additional details in supplementary section 3. The agreement between the fit and the data indicates that the dynamics of the order parameter are indeed governed by a fluence and time dependent double-well potential. Moreover, it shows that using our pump pulse we can control the shape of the double-well, which triggers switching the sign of \PFE~on picosecond timescales. The shape of the double-well potential is shown for selected time delays in Figure \ref{Figure 4}(c-e) along with the position and trajectory of  \PFE~marked by an orange circle and a black arrow.

\par
Our measurements reveal a separation between the transition temperature $T_c$ and the higher temperature $T_0$ obtained by extrapolating the soft mode stiffening. In a purely displacive ferroelectric, the polar phonon instability directly drives the transition, so $T_c=T_0$. Conversely, in the order-disorder limit, criticality is dominated by slow relaxational polarization dynamics and the THz phonon frequency softens only weakly, if at all \cite{huber1972critical}. SnTe exhibits a clean second order transition in SHG while still showing pronounced softening of the ferroelectric mode, pointing to a mixed transition in which
different frequency sectors renormalize differently. This interpretation is supported by recent reports of Rashba splitting persisting to temperatures well above $T_c$ \cite{Chassot2024NanoLett}, intriguingly, its temperature dependence is approximately mean-field like up to $\sim$100K, close to the $T_0$ inferred here, before saturating at higher temperature. Consistently, scanning-probe measurements have identified nanoscale ferroelectric domains above $T_c$ \cite{aggarwal2016local}.
\par 

Even more striking than the separation of \Tc~and $T_0$~is the non adiabatic separation between the out of equilibrium dynamics of the phonon coordinate $Q$ (probed by \trR) and \PFE~(probed by SHG). Our results show that the two do not share the same potential energy landscape at high fluences. We believe a similar separation will be detectable in other ferroelectrics where a hybrid ionic-electronic character of the soft mode is expected.
\par


Phenomenologically, renormalization of the quadratic stiffness, i.e., $\alpha(T)\rightarrow \alpha(T)+\Delta\alpha$, will renormalize $T_c$ and $T_0$. The fact that $T_0>T_c$ implies that the renormalization of the quadratic stiffness is different for the ferroelectric mode softening and the formation of long range order \PFE. While the critical temperature $T_c$ is set by the long, macroscopic length scale of the polar response and dipole-dipole interactions (small $q$), $T_0$~obtained from the Cochran plot reflects the stiffness of the soft mode in which is determined by intra-unit cell potentials (large $q$). If the same renormalization acted equally for both length scales, it would shift both scales in tandem, therefore, the observed separation requires a scale dependent relormalization: $\Delta\alpha_{P_{FE}}\neq \Delta\alpha_{phonon}$.

This viewpoint also provides a natural link to the nonequilibrium dynamics: in our quench model, the pump enters through a time dependent parameter $\eta(t)$ that renormalizes the curvature of the double-well potential for $P_{\mathrm{FE}}$. The key empirical observation is that, at high fluence, the order parameter extracted from phase sensitive RA-SHG develops slow, strongly nonlinear dynamics, while the coherent oscillation frequency observed in $\Delta R/R$ remains comparatively unchanged. This is precisely the behavior expected if the photoexcitation ($\eta(t)$) primarily modifies the long range stiffness relevant to the polarization sector, without equivalently renormalizing the stiffness on the unit-cell lengthscale that governs the small amplitude coherent oscillation.

Screening by free charge carriers strongly suppresses $T_c$ in SnTe \cite{SugaiMurase1982}, and at a sufficiently high carrier density, the ferroelectric transition can be suppressed altogether. We therefore suggest that screening provides the microscopic origin for the stiffness renormalization expressed by $\Delta\alpha$ and $\eta(t)$ and naturally yields different renormalization for intra unit cell phenomena such as the phonon oscillations and long length scale phenomena such as \PFE.

In summary, we combine phase sensitive time resolved SHG and ultrafast pump-probe reflectivity to independently track the ferroelectric polarization $P_{\mathrm{FE}}(t)$ and the coherent lattice coordinate $\Delta Q(t)$ of the soft mode in SnTe. Photoexcited carriers transiently reshape the double-well free energy landscape and enable ultrafast polarization switching. At high excitation density, $P_{\mathrm{FE}}(t)$ develops slow, strongly anharmonic dynamics, while the coherent oscillation frequency observed in \trR~ remains nearly unchanged, revealing a breakdown of adiabatic locking between electronic polarization and ionic motion. Together with the separation between $T_c$ and the Cochran extrapolation temperature $T_0$, our results point to a scale dependent renormalization of the polar stiffness and highlight the hybrid electronic and ionic character of the ferroelectric mode. Additionally, this work demonstrates carrier driven landscape engineering as a route to control ferroelectric order on ultrafast timescales, and that phase resolved SHG can directly determine polarization sign and distinguish domain orientations.

\begin{addendum}
\item [Methods] SnTe single crystals were grown using the vapor transport technique without the aid of a transport agent, similar to the procedure described in Ref \cite{kannaujiya2020growth}. Stoichiometric amounts of Sn(99.995\% Sigma Aldrich) and Te(99.999\% Thermoscientific) were weighed and thoroughly mixed inside an alumina crucible. The starting materials were sealed under vacuum in a 20cm long quartz ampule which was placed in a gradient tube furnace. The starting materials were positioned in the hot zone of the furnace. The source and growth zones of the furnace were ramped to 797\textdegree C and 747\textdegree C, respectively, at a rate of 25\textdegree C/hour. The synthesis took place for 1 week after which large single crystals were found in the cold zone of the tube. See structural characterization in Ref \cite{scharf2025coherent}. 
\par
Additional experimental details are given in supplementary section 1. Error bars throughout the manuscript correspond to a single standard deviation extracted from the corresponding fit.
\item A.R. acknowledges support from the Zuckerman Foundation, and the Israel Science Foundation (Grant No. 592/25). This research was supported by the Ministry of Innovation, Science and Technology, Israel. This research was supported by the Ministry of Energy, Israel. Gili Scharf acknowledges support from the Israeli Clore fellowship and Nova Ltd. We thank Hadas Soifer, Eran Sela, Dominik Juraschek, Jonathan Curtis, and Tobias Holder for fruitful discussions.

\item [Data availability] The data that support the findings of this study are available from the corresponding author upon reasonable request. 

\item [Author contribution] All authors contributed to the production of this manuscript.
\item[Competing Interests] The authors declare no competing financial interests.
\item[Correspondence] Correspondence and requests for materials
should be addressed to Alon Ron~(email: alonron@tauex.tau.ac.il).
\end{addendum}

\bibliography{SnTe111.bib}

@article{Cochran1960,
  author  = {Cochran, W.},
  title   = {Crystal stability and the theory of ferroelectricity},
  journal = {Advances in Physics},
  volume  = {9},
  pages   = {387--423},
  year    = {1960}
}

@article{Pawley1966,
  author  = {Pawley, G. S. and Cochran, W. and Cowley, R. A. and Dolling, G.},
  title   = {Anharmonic effects in the lattice dynamics of SnTe},
  journal = {Physical Review Letters},
  volume  = {17},
  pages   = {753--756},
  year    = {1966}
}

@article{Littlewood1980,
  author  = {Littlewood, P. B.},
  title   = {Theory of ferroelectric phase transitions in IV-VI semiconductors},
  journal = {Journal of Physics C: Solid State Physics},
  volume  = {13},
  pages   = {4855--4873},
  year    = {1980}
}

@article{SugaiMurase1982,
  author  = {Sugai, S. and Murase, K.},
  title   = {Soft-mode behavior in SnTe studied by Raman scattering},
  journal = {Physical Review B},
  volume  = {25},
  pages   = {2418--2426},
  year    = {1982}
}

@article{Polking2012,
  author  = {Polking, M. J. and Han, M.-G. and Yourdkhani, A. and others},
  title   = {Ferroelectric order in individual SnTe nanoplates},
  journal = {Physical Review Letters},
  volume  = {109},
  pages   = {247601},
  year    = {2012}
}

@article{seyler2022direct,
  title={Direct visualization and control of antiferromagnetic domains and spin reorientation in a parent cuprate},
  author={Seyler, KL and Ron, A and Van Beveren, D and Rotundu, CR and Lee, YS and Hsieh, D},
  journal={Physical Review B},
  volume={106},
  number={14},
  pages={L140403},
  year={2022},
  publisher={APS}
}

@article{Mitrofanov2014PRB,
  author  = {Mitrofanov, K. V. and Kolobov, A. V. and Fons, P. and Krbal, M. and Shintani, T. and Tominaga, J. and Uruga, T.},
  title   = {Local structure of the {SnTe} topological crystalline insulator: Rhombohedral distortions emerging from the rocksalt phase},
  journal = {Physical Review B},
  year    = {2014},
  volume  = {90},
  number  = {13},
  pages   = {134101},
  doi     = {10.1103/PhysRevB.90.134101}
}

@article{Fornasini2021JPCM,
  author  = {Fornasini, P. and Grisenti, R. and Dapiaggi, M. and Agostini, G.},
  title   = {Local structural distortions in {SnTe} investigated by {EXAFS}},
  journal = {Journal of Physics: Condensed Matter},
  year    = {2021},
  volume  = {33},
  number  = {29},
  pages   = {295404},
  doi     = {10.1088/1361-648X/ac0082}
}

@article{Aggarwal2016JMat,
  author  = {Aggarwal, Leena and Banik, Ananya and Anand, Shashwat and Waghmare, Umesh V. and Biswas, Kanishka and Sheet, Goutam},
  title   = {Local ferroelectricity in thermoelectric {SnTe} above room temperature driven by competing phonon instabilities and soft resonant bonding},
  journal = {Journal of Materiomics},
  year    = {2016},
  volume  = {2},
  number  = {2},
  pages   = {196--202},
  doi     = {10.1016/j.jmat.2016.04.001}
}

@article{Chassot2024NanoLett,
  author  = {Chassot, Fr\'ed\'eric and Pulkkinen, Aki and Kremer, Geoffroy and Zakusylo, Tetiana and Krizman, Gauthier and Hajlaoui, Mahdi and Dil, J. Hugo and Krempask\'y, Juraj and Min\'ar, J\'an and Springholz, Gunther and Monney, Claude},
  title   = {Persistence of Structural Distortion and Bulk Band {Rashba} Splitting in {SnTe} above Its Ferroelectric Critical Temperature},
  journal = {Nano Letters},
  year    = {2024},
  volume  = {24},
  number  = {1},
  pages   = {82--88},
  doi     = {10.1021/acs.nanolett.3c03280}
}

@article{kannaujiya2020growth,
  title={Growth and characterizations of tin telluride (SnTe) single crystals},
  author={Kannaujiya, Rohitkumar M and Khimani, Ankurkumar J and Chaki, Sunil H and Chauhan, Sanjaysinh M and Hirpara, Anilkumar B and Deshpande, MP},
  journal={The European Physical Journal Plus},
  volume={135},
  pages={1--12},
  year={2020},
  publisher={Springer}
}

@article{Gallego:lk5043,
author = "Gallego, Samuel V. and Etxebarria, Jesus and Elcoro, Luis and Tasci, Emre S. and Perez-Mato, J. Manuel",
title = "{Automatic calculation of symmetry-adapted tensors in magnetic and non-magnetic materials: a new tool of the Bilbao Crystallographic Server}",
journal = "Acta Crystallographica Section A",
year = "2019",
volume = "75",
number = "3",
pages = "438--447",
month = "May",
doi = {10.1107/S2053273319001748},
url = {https://doi.org/10.1107/S2053273319001748},
}

@article{Aroyo2006,
url = {https://doi.org/10.1524/zkri.2006.221.1.15},
title = {Bilbao Crystallographic Server: I. Databases and crystallographic computing programs},
title = {},
author = {Mois Ilia Aroyo and Juan Manuel Perez-Mato and Cesar Capillas and Eli Kroumova and Svetoslav Ivantchev and Gotzon Madariaga and Asen Kirov and Hans Wondratschek},
pages = {15--27},
volume = {221},
number = {1},
journal = {Zeitschrift für Kristallographie - Crystalline Materials},
doi = {doi:10.1524/zkri.2006.221.1.15},
year = {2006},
lastchecked = {2026-01-26}
}

@article{Aroyo2011,
  author = {Aroyo, M. I. and Perez-Mato, J. M. and Orobengoa, D. and Tasci, E. and de la Flor, G. and Kirov, A.},
  title = {Crystallography online: {B}ilbao {C}rystallographic {S}erver},
  journal = {Bulgarian Chemical Communications},
  volume = {43},
  number = {2},
  pages = {183--197},
  year = {2011}
}

@article{Aroyo:xo5013,
author = "Aroyo, Mois I. and Kirov, Asen and Capillas, Cesar and Perez-Mato, J. M. and Wondratschek, Hans",
title = "{Bilbao Crystallographic Server. II. Representations of crystallographic point groups and space groups}",
journal = "Acta Crystallographica Section A",
year = "2006",
volume = "62",
number = "2",
pages = "115--128",
month = "Mar",
doi = {10.1107/S0108767305040286},
url = {https://doi.org/10.1107/S0108767305040286},
}

@article{ron2019dimensional,
  title={Dimensional crossover in a layered ferromagnet detected by spin correlation driven distortions},
  author={Ron, A and Zoghlin, E and Balents, L and Wilson, SD and Hsieh, D},
  journal={Nature communications},
  volume={10},
  number={1},
  pages={1654},
  year={2019},
  publisher={Nature Publishing Group UK London}
}

@article{sugai1977carrier,
  title={Carrier density dependence of soft TO-phonon in SnTe by Raman scattering},
  author={Sugai, S and Murase, K and Katayama, S and Takaoka, S and Nishi, S and Kawamura, H},
  journal={Solid State Communications},
  volume={24},
  number={5},
  pages={407--409},
  year={1977},
  publisher={Elsevier}
}

@article{zeiger1992theory,
  title={Theory for displacive excitation of coherent phonons},
  author={Zeiger, HJ and Vidal, J and Cheng, TK and Ippen, EP and Dresselhaus, G and Dresselhaus, MS},
  journal={Physical Review B},
  volume={45},
  number={2},
  pages={768},
  year={1992},
  publisher={APS}
}

@article{huber2014coherent,
  title={Coherent structural dynamics of a prototypical charge-density-wave-to-metal transition},
  author={Huber, Tim and Mariager, Simon O and Ferrer, Andres and Sch{\"a}fer, Hanjo and Johnson, Jeremy A and Gr{\"u}bel, Sebastian and L{\"u}bcke, Andre and Huber, Lucas and Kubacka, Teresa and Dornes, Christian and others},
  journal={Physical review letters},
  volume={113},
  number={2},
  pages={026401},
  year={2014},
  publisher={APS}
}

@article{yusupov2010coherent,
  title={Coherent dynamics of macroscopic electronic order through a symmetry breaking transition},
  author={Yusupov, Roman and Mertelj, Tomaz and Kabanov, Viktor V and Brazovskii, Serguei and Kusar, Primoz and Chu, Jiun-Haw and Fisher, Ian R and Mihailovic, Dragan},
  journal={Nature Physics},
  volume={6},
  number={9},
  pages={681--684},
  year={2010},
  publisher={Nature Publishing Group UK London}
}

@article{schaefer2014collective,
  title={Collective modes in quasi-one-dimensional charge-density wave systems probed by femtosecond time-resolved optical studies},
  author={Schaefer, Hanjo and Kabanov, Viktor V and Demsar, Jure},
  journal={Physical Review B},
  volume={89},
  number={4},
  pages={045106},
  year={2014},
  publisher={APS}
}

@article{huber1972critical,
  title={Critical Dynamics of the Order-Disorder Transformation in Ferroelectrics},
  author={Huber, DL},
  journal={Physical Review B},
  volume={6},
  number={9},
  pages={3379},
  year={1972},
  publisher={APS}
}

@article{aggarwal2016local,
  title={Local ferroelectricity in thermoelectric SnTe above room temperature driven by competing phonon instabilities and soft resonant bonding},
  author={Aggarwal, Leena and Banik, Ananya and Anand, Shashwat and Waghmare, Umesh V and Biswas, Kanishka and Sheet, Goutam},
  journal={Journal of Materiomics},
  volume={2},
  number={2},
  pages={196--202},
  year={2016},
  publisher={Elsevier}
}

@article{scharf2025coherent,
  title={Coherent control through phonon anharmonicity},
  author={Scharf, Gili and Hasharoni, Tomer and Donval, Lara and Gur, Leah Ben and Ron, Alon},
  journal={arXiv preprint arXiv:2503.10757},
  year={2025}
}

@article{BornOppenheimer1927,
  author  = {Born, M. and Oppenheimer, R.},
  title   = {Zur Quantentheorie der Molekeln},
  journal = {Annalen der Physik},
  volume  = {389},
  number  = {20},
  pages   = {457--484},
  year    = {1927},
  doi     = {10.1002/andp.19273892002}
}

@article{HohenbergKohn1964,
  author  = {Hohenberg, P. and Kohn, W.},
  title   = {Inhomogeneous Electron Gas},
  journal = {Physical Review},
  volume  = {136},
  pages   = {B864--B871},
  year    = {1964},
  doi     = {10.1103/PhysRev.136.B864}
}

@article{KohnSham1965,
  author  = {Kohn, W. and Sham, L. J.},
  title   = {Self-Consistent Equations Including Exchange and Correlation Effects},
  journal = {Physical Review},
  volume  = {140},
  pages   = {A1133--A1138},
  year    = {1965},
  doi     = {10.1103/PhysRev.140.A1133}
}

@article{GonzeLee1997,
  author  = {Gonze, X. and Lee, C.},
  title   = {Dynamical matrices, Born effective charges, dielectric permittivity tensors, and interatomic force constants from density-functional perturbation theory},
  journal = {Physical Review B},
  volume  = {55},
  pages   = {10355--10368},
  year    = {1997},
  doi     = {10.1103/PhysRevB.55.10355}
}

@article{Baroni2001,
  author  = {Baroni, S. and de Gironcoli, S. and Dal Corso, A. and Giannozzi, P.},
  title   = {Phonons and related crystal properties from density-functional perturbation theory},
  journal = {Reviews of Modern Physics},
  volume  = {73},
  pages   = {515--562},
  year    = {2001},
  doi     = {10.1103/RevModPhys.73.515}
}

@article{iizumi1975phase,
  title={Phase transition in SnTe with low carrier concentration},
  author={Iizumi, Masashi and Hamaguchi, Yoshikazu and F. Komatsubara, Kiichi and Kato, Yoshiki},
  journal={Journal of the Physical Society of Japan},
  volume={38},
  number={2},
  pages={443--449},
  year={1975},
  publisher={The Physical Society of Japan}
}

@article{o2017inelastic,
  title={Inelastic x-ray investigation of the ferroelectric transition in SnTe},
  author={O'Neill, Christopher D and Sokolov, Dmitry A and Hermann, Andreas and Bossak, Alexei and Stock, Christopher and Huxley, Andrew D},
  journal={Physical Review B},
  volume={95},
  number={14},
  pages={144101},
  year={2017},
  publisher={APS}
}

@article{LazzeriMauri2006,
  title        = {Nonadiabatic Kohn Anomaly in a Doped Graphene Monolayer},
  author       = {Lazzeri, Michele and Mauri, Francesco},
  journal      = {Phys. Rev. Lett.},
  volume       = {97},
  number       = {26},
  pages        = {266407},
  year         = {2006},
  month        = dec,
  publisher    = {American Physical Society},
  doi          = {10.1103/PhysRevLett.97.266407}
}

@article{Saitta2008,
  title        = {Giant Nonadiabatic Effects in Layer Metals: Raman Spectra of Intercalated Graphite Explained},
  author       = {Saitta, A. Marco and Lazzeri, Michele and Calandra, Matteo and Mauri, Francesco},
  journal      = {Phys. Rev. Lett.},
  volume       = {100},
  number       = {22},
  pages        = {226401},
  year         = {2008},
  month        = jun,
  publisher    = {American Physical Society},
  doi          = {10.1103/PhysRevLett.100.226401}
}

@article{Hoang2025,
  title        = {Ultrafast decoupling of polarization and strain in ferroelectric {BaTiO$_3$}},
  author       = {Hoang, Le Phuong and Pesquera, David and Hinsley, Gerard N. and Carley, Robert and Mercadier, Laurent and Teichmann, Martin and Unterleutner, Elena Martina and Knez, Daniel and Dienstleder, Martina and Ganguly, Saptam and Asmara, Teguh Citra and Merzoni, Giacomo and Parchenko, Sergii and Schlappa, Justine and Yin, Zhong and Caicedo Roque, Jose Manuel and Santiso, Jose and Spasojevic, Irena and Carinan, Cammille and Lee, Tien-Lin and Rossnagel, Kai and Zegenhagen, J{\"o}rg and Catalan, Gustau and Vartanyants, Ivan A. and Scherz, Andreas and Mercurio, Giuseppe},
  journal      = {Nature Communications},
  volume       = {16},
  pages        = {7966},
  year         = {2025},
  doi          = {10.1038/s41467-025-63045-6}
}

@article{Petzelt1984,
  title = {Central-peak--soft-mode coupling in ferroelectric Gd$_2$(MoO$_4$)$_3$},
  author = {Petzelt, J. and Smutny, F. and Katkanant, V. and Ullman, F. G. and Hardy, J. R. and Volkov, A. A. and Kozlov, G. V. and Lebedev, S. P.},
  journal = {Phys. Rev. B},
  volume = {30},
  pages = {5172},
  year = {1984},
  doi = {10.1103/PhysRevB.30.5172}
}

\newpage

\setcounter{figure}{0}

\vspace{1cm}
\begin{center}
\begin{spacing}{1.25} 
    \textbf{\large Supplemental Material for: "Non adiabatic dynamics of the ferroelectric soft mode"}
\end{spacing}
\end{center}



\makeatletter
\let\saved@includegraphics\includegraphics
\AtBeginDocument{\let\includegraphics\saved@includegraphics}
\renewenvironment{figure}{\@float{figure}}
{\end@float}
\makeatother

\renewcommand{\thefigure}{S\arabic{figure}}





\maketitle



\section{Optical setup}
1030nm pulses were generated by a regenerative amplifier at a repetition rate of 100kHz, these were used to pump a pair of optical parametric amplifiers (OPAs) seeded by the same white light crystal. The signal beam from one OPA was set to 900nm and used as a probe, whereas the idler beam from the second OPA was set to 1400nm and used as the pump. The pulse durations of the two beams were 99.4[fs] and 137.8[fs], respectively. The RA-SHG measurements were taken in the polarizer analyzer geometry. The probe beam polarization was rotated using a $\lambda/2$~waveplate and the $2\omega$~signal at 450nm collected from the sample was passed through a linear polarizer and detected by a photomultiplier tube with the aid of three band-pass filters centered at 450nm. For the measurements, the relative angles of the polarizer/waveplate were set such that the polarizations are either parallel or perpendicular to one another, depending on the measurement configuration. For the pump-probe measurements, the basic repetition rate of the laser was set to 50kHz and a mechanical chopper was used to modulate the pump beam at 5kHz. For these measurements, the probe beam was collected by a photodiode and the pump fluence was tuned using a combination of a linear polarizer and a $\lambda/2$~waveplate. The temperature was controlled using a sample in a vacuum-closed cycle optical cryostat.
\begin{figure}[!ht]

    \centering
    \includegraphics[width=1\textwidth]{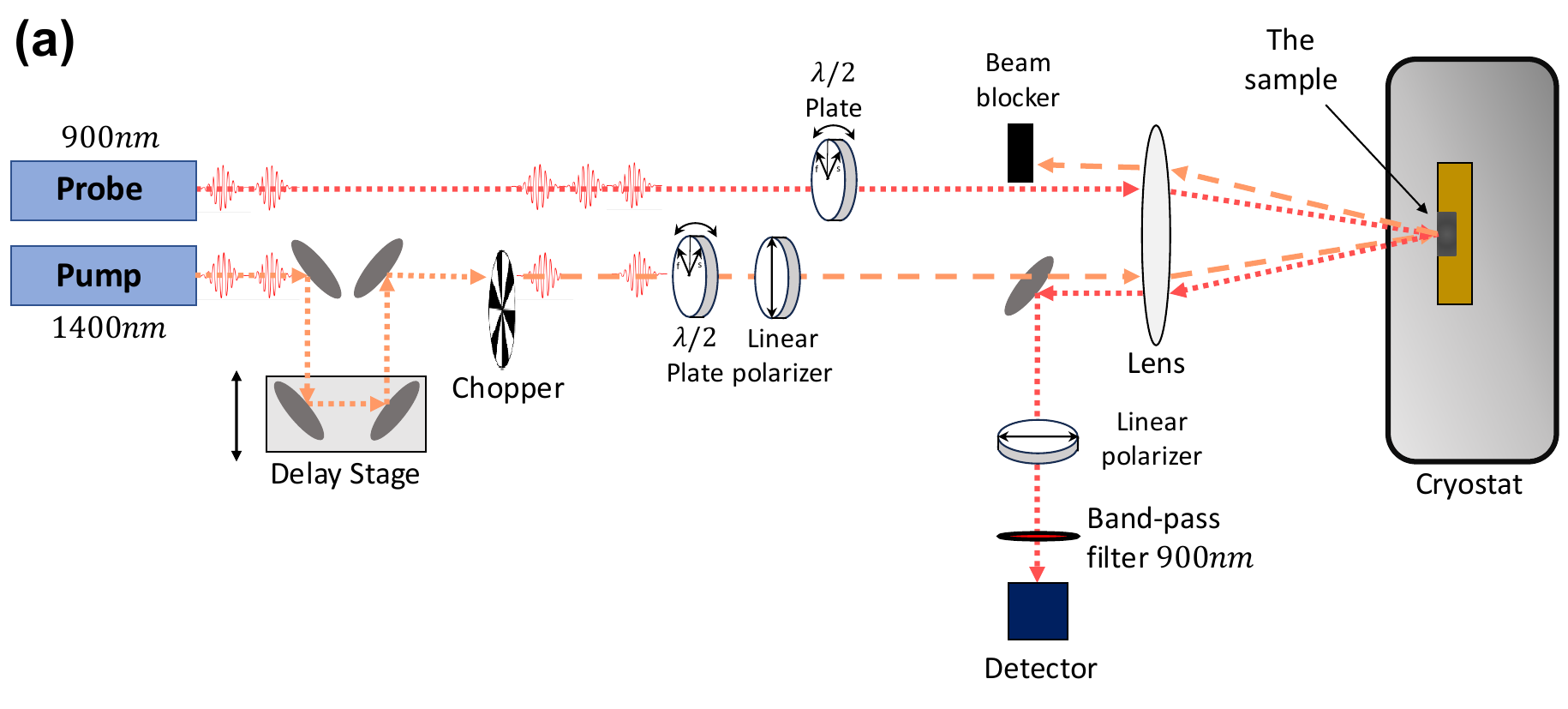}
    \includegraphics[width=1\textwidth]{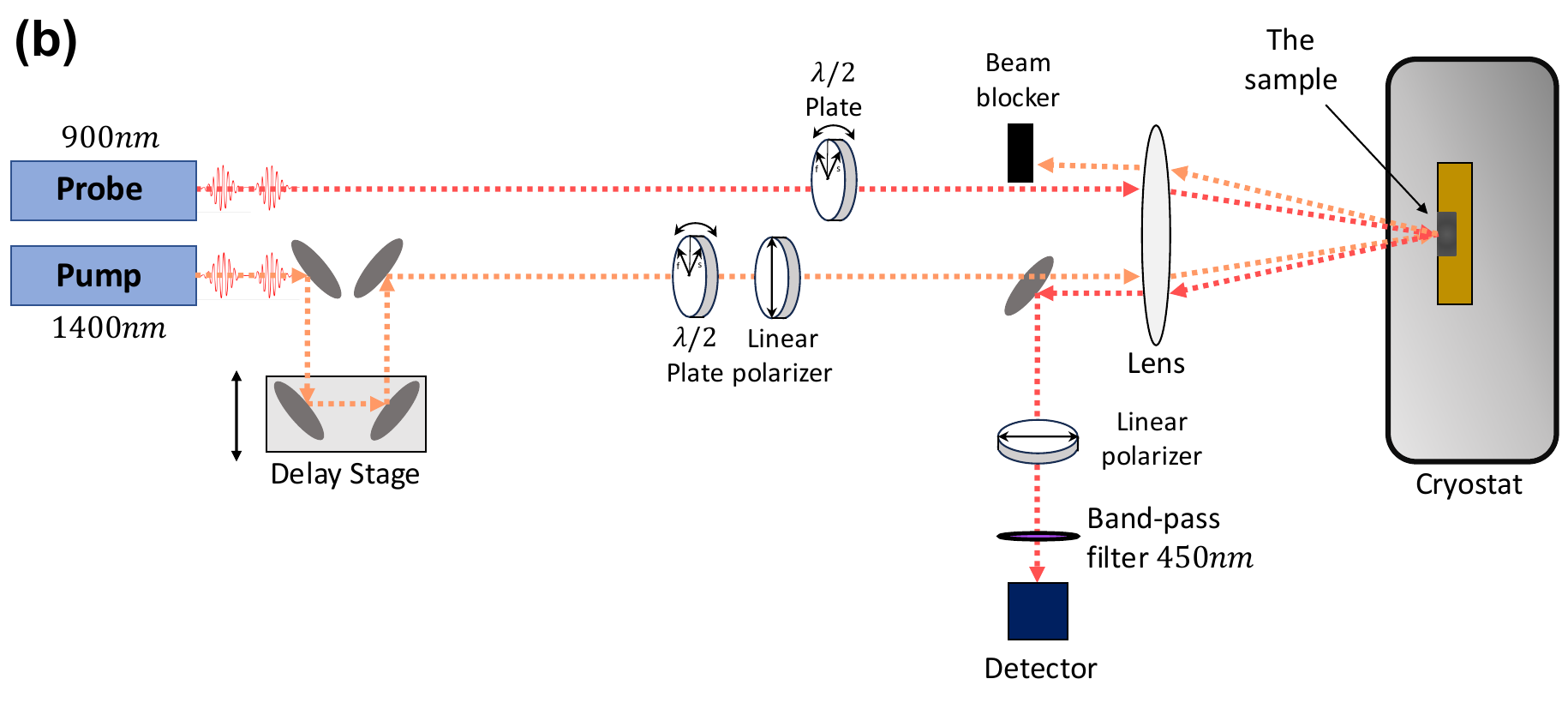}
    \caption{Schematic of the optical setup used in this work. (a) The optical setup used for the pump-probe measurements, and (b) the optical setup for the SHG measurements. Unlabeled components are silver mirrors.}
    \label{SHG setup figure}
\end{figure}

\clearpage

\section{The SHG tensors and response for SnTe (111) surface}
As mentioned in the main text, the point group symmetry corresponding to the SnTe (111) surface is $3m$, where $\hat e_3 \parallel (111)$ \cite{Aroyo2006,Aroyo:xo5013,Aroyo2011}. The SHG tensor for this point group is \cite{Aroyo2006,Aroyo:xo5013,Aroyo2011,Gallego:lk5043}:
\begin{equation}
    \chi_{\mathrm{SHG}-3m}(2\omega;\omega,\omega)=
        \begin{pmatrix}
        0 & 0 & 0 & 0 & \chi_{15} & \chi_{16} \\
        \chi_{16} & -\chi_{16} & 0 & \chi_{15} & 0 & 0 \\
        \chi_{31} & \chi_{31} & \chi_{33} & 0 & 0 & 0
        \end{pmatrix}
\end{equation}
where the 2nd index is in Voigt notation.
In our experiment, since the probe is (approximately) in normal incidence to the surface, the incident and reflected electric fields have only in-plane components. Plugging this into $P_i(2\omega)=\chi_{ijk}^{(2)}E_j(\omega)E_k(\omega)$~results in:
\begin{align}
    &P_{SHG-3m}^{\parallel}(2\omega, \phi) = \chi_{16}E^2(\omega)\sin(3\phi) \\
    &P_{SHG-3m}^{\perp}(2\omega, \phi) = \chi_{16}E^2(\omega)\cos(3\phi) 
\end{align}
and the following intensities which are measured in the experiment:
\begin{align}
    &I_{SHG-3m}^{\parallel}=\Big|P_{SHG}^{\parallel}(2\omega, \phi)\Big|^2 = \frac{1}{2}\chi_{16}^2E^4(\omega)\big(1-\cos(6\phi)\big) \\
    &I_{SHG-3m}^{\perp}=\Big|P_{SHG}^{\perp}(2\omega, \phi)\Big|^2 = \frac{1}{2}\chi_{16}^2E^4(\omega)\big(1+\cos(6\phi)\big) 
\end{align}

In the ferroelectric state, the $3m$ point group symmetry is preserved, but on an equivalent \{111\} plane different than the naturally grown (111) orientation of our sample. This lowers the point group symmetry to $m$ \cite{Aroyo2006,Aroyo:xo5013,Aroyo2011} which has the following SHG tensor \cite{Aroyo2006,Aroyo:xo5013,Aroyo2011,Gallego:lk5043}:

\begin{equation}
    \chi_{\mathrm{SHG}-m}(2\omega;\omega,\omega)=
        \begin{pmatrix}
        \chi'_{11} & \chi'_{12} & \chi'_{13} & 0 & \chi'_{15} & 0 \\
        0 & 0 & 0 & \chi'_{24} & 0 & \chi'_{26} \\
        \chi'_{31} & \chi'_{31} & \chi'_{33} & 0 & \chi'_{35} & 0
        \end{pmatrix}
\end{equation}
where the 2nd index is in Voigt notation.
This results in the following form for the $2\omega$~polarization:
\begin{align}
    &P_{SHG-m}^{\parallel}(2\omega, \phi) = \frac{1}{4}E^2(\omega)\big[(\chi'_{11}-\chi'_{12}-2\chi'_{26})\cos(3\phi)+(3\chi'_{11}+\chi'_{12}+2\chi'_{26})\cos(\phi)\big] \\
    &P_{SHG-m}^{\perp}(2\omega, \phi) = \frac{1}{4}E^2(\omega)\big[(\chi'_{12}-\chi'_{11}+2\chi'_{26})\sin(3\phi)-(3\chi'_{12}+\chi'_{11}-2\chi'_{26})\sin(\phi)\big] 
\end{align}
And the following experimentally accessible intensities:
\begin{align}
I_{\parallel}(\phi)
&=\Big|P_{SHG-m}^{\parallel}(2\omega,\phi)\Big|^2 \nonumber\\
&=\frac{1}{16}E^4(\omega)
\left[
\begin{aligned}
&(\chi'_{11}-\chi'_{12}-2\chi'_{26})\cos(3\phi)\\
&\qquad +(3\chi'_{11}+\chi'_{12}+2\chi'_{26})\cos(\phi)
\end{aligned}
\right]^2 \nonumber\\
&=\left(\chi^\parallel_{3\mathrm{fold}}\cos(3\phi)+\chi^\parallel_{2\mathrm{fold}}\cos(\phi)\right)^2 ,
\\[6pt]
I_{\perp}(\phi)
&=\Big|P_{SHG-m}^{\perp}(2\omega,\phi)\Big|^2 \nonumber\\
&=\frac{1}{16}E^4(\omega)
\left[
\begin{aligned}
&(\chi'_{12}-\chi'_{11}+2\chi'_{26})\sin(3\phi)\\
&\qquad -(3\chi'_{12}+\chi'_{11}-2\chi'_{26})\sin(\phi)
\end{aligned}
\right]^2 \nonumber\\
&=\left(\chi^\perp_{3\mathrm{fold}}\sin(3\phi)-\chi^\perp_{2\mathrm{fold}}\sin(\phi)\right)^2
\end{align}
where:
\begin{align}
\chi^\parallel_{3\mathrm{fold}} &= \frac{E^2(\omega)}{4}\left(\chi'_{11}-\chi'_{12}-2\chi'_{26}\right),\\
\chi^\parallel_{2\mathrm{fold}} &= \frac{E^2(\omega)}{4}\left(3\chi'_{11}+\chi'_{12}+2\chi'_{26}\right),\\
\chi^\perp_{3\mathrm{fold}} &= \frac{E^2(\omega)}{4}\left(\chi'_{12}-\chi'_{11}+2\chi'_{26}\right),\\
\chi^\perp_{2\mathrm{fold}} &= \frac{E^2(\omega)}{4}\left(3\chi'_{12}+\chi'_{11}-2\chi'_{26}\right).
\end{align}
are the coefficients used for fitting our data.

\clearpage

\section{Fitting procedure and results}
In the main text, the fitting procedure of \PFE~as a function of time delay to the following equation of motion is described:
\begin{equation}
    \frac{1}{\omega^2}\frac{\partial ^2}{\partial t^2}P_{FE}-(\eta(t)-1)P_{FE}+P_{FE}^3+\frac{2\gamma}{\omega^2}\frac{\partial}{\partial t}P_{FE}=0
\end{equation}
 Where $\eta(t)=\eta_0\cdot g(t)\cdot \mathrm{exp}(-t/\tau)+\eta_\infty$~and $g(t)=\mathrm{erf}(\sigma t)$. The initial conditions for solving the equation are that $P_{FE}(t=0)=-1$~and that $\frac{dP_{FE}(t=0)}{dt}=0$, which were set for the solvers initial conditions. To satisfy the position's initial condition, the data was normalized to its value under equilibrium conditions and set (arbitrarily) to have a negative sign compatible with one of the two identical double wells. 

The fitting parameters of this model to the data shown in Figure 4 of the main text and below are: $\eta_\infty = 0.83$, $\tau = 0.20 [ps]$, $\sigma = 0.52 [ps]$, $\gamma = 0.35 [THz] $.  The angular frequency $\omega$~which is the stiffness of the mode, was fixed (not a fitting parameter) at its experimental value from the pump probe measurements at 5K:  $2\pi\cdot f =2\pi \cdot 0.9105 [THz]$. The fit results (identical to the main text Figure 4(a) as well as the result for $\eta(t)$~are shown below. The first oscillatory peak dynamics seem to be dominated by the shape of $\eta(t)$~and specifically by the time-scale $1/\sigma$. This is because $\sigma$~represents the timescale it takes the double-well structure to respond to the pump. In contrast, since $\eta(t)$~has relaxed to its asymptomatic value after $\simeq2[ps]$, the late time dynamics and in particular the second and third oscillations are dominated by the dynamics in the double-well structure. 
\begin{figure}
    \centering
    \includegraphics[width=1\linewidth]{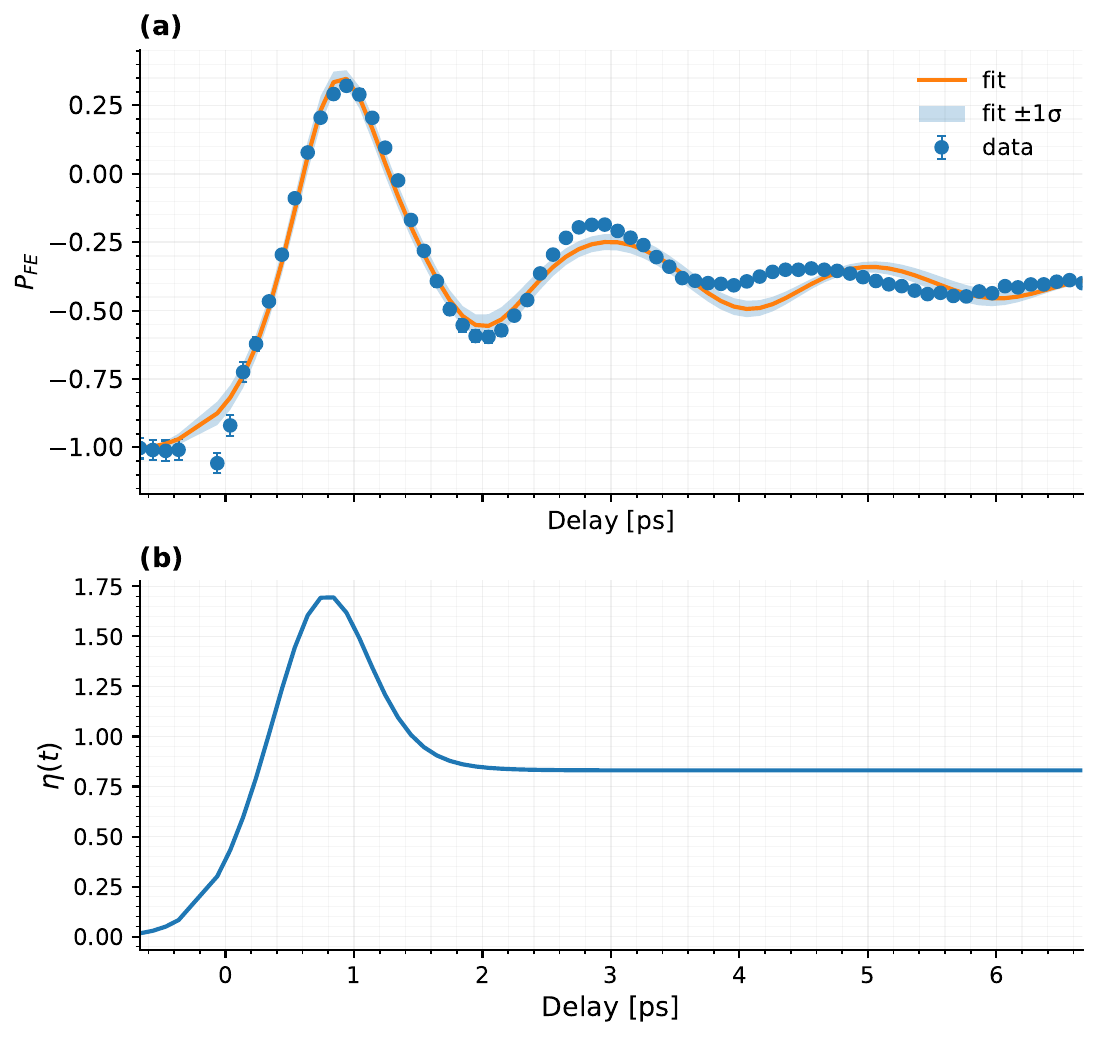}
    \caption{(a) Fit of \PFE~to the model described in the text. Blue points are the data and the orange curve is the fit. (b) $\eta(t)$~as a function of time delay as extracted from the fit.}
    \label{fig:placeholder}
\end{figure}




\end{document}